# Proposal and demonstration of germanium-on-silicon lock-in pixels for indirect time-of-flight based three-dimensional sensing

**N. Na, S.-L. Cheng, H.-D. Liu, M.-J. Yang, C.-Y. Chen, K.-C. Chu, H.-W. Chen, Y.-T. Chou, C.-T. Lin, W.-H. Liu, C.-F. Liang, C.-L. Chen, S.-W. Chu, B.-J. Chen, Y.-F. Lyu, and S.-L. Chen**

*Artilux Inc., 8F-1 No. 6 Taiyuan 1st Rd., Zhubei City, Hsinchu County, Taiwan 30288, ROC*

**Abstract:** We propose the use of germanium-on-silicon technology for indirect time-of-flight based three-dimensional sensing, and demonstrate a novel lock-in pixel featuring high quantum efficiency and large frequency bandwidth. Compared to silicon pixels, germanium-on-silicon pixels simultaneously maintain a high quantum efficiency and a high demodulation contrast deep into GHz frequency regime, which enable consistently superior depth accuracy in both indoor and outdoor scenarios. Physical model, numerical simulation, device fabrication and characterization, system performance comparison, and laser safety analysis are presented. Our work paves a new path to high-performance time-of-flight rangers and imagers, as well as potential adoption of lasers operated at a longer near infrared wavelength that falls outside of the operation window of silicon pixels.

## 1. Introduction

The construction of three-dimensional (3D) images is vital to a variety of applications such as hand tracking, facial recognition, 3D object scanning/printing, simultaneous localization and mapping (SLAM) for indoor/outdoor navigations, surveillance, and light detection and ranging (LiDAR) for autonomous vehicles, to name a few. There are a least four categories of techniques that are well studied in the literature, i.e., passive/active stereovision [1], structured light imaging [2], direct time-of-flight (TOF) sensing via single photon avalanche photodiode (SPAD) and time-to-digital converter (TDC) [3], and indirect TOF sensing via lock-in pixel [4]. In particular, the indirect TOF system has been a popular choice due to its direct acquisition of a depth information without additional computational algorithms, simple laser light projector without complex optical modules for spatial coded patterns, and low power consumption with respect to pixel number scaling. The principle of the indirect TOF system is essentially based on a homodyne detection operated at a radio frequency (RF) $f_m$, in which the roles of a local oscillator/mixer/detector are replaced by a lock-in pixel, to demodulate the transmitted laser light that is initially modulated at $f_m$, then reflected from a 3D object, and finally received by the lock-in pixel. Consequently, the phase of the laser light can be extracted by detecting the low-frequency signal amplitudes, which indirectly maps the distance that the laser light traverses.

Various lock-in pixels have been analyzed and implemented using silicon (Si) technology, in which the demodulations are achieved through capacitive switching devices, such as charge-coupled device (CCD) [5], complementary-metal-oxide-semiconductor (CMOS) transfer gate [6], demodulated photogate [7-8], and pinned photodiode with transfer gate [9], or through resistive switching devices, such as current assisted photonic demodulator (CAPD) [10-13]. To achieve a high signal-to-noise ratio and therefore a small depth error in an indirect TOF measurement, it is desired to increase the quantum efficiency and the frequency bandwidth of a lock-in pixel. With the recent trend to shift the near-infrared (NIR) operation wavelength from 850 nm to 940 nm to take advantage of the solar spectrum dip at 940 nm (i.e., operating at a low ambient light noise condition to expand the 3D sensing applications from indoor to outdoor scenario), the quantum efficiency degrades further as Si features weaker absorption at longer NIR wavelengths. Limited by the efficiency-bandwidth-product, while it is possible to obtain a higher quantum efficiency by simply having a thicker Si absorption layer, the resultant lower frequency bandwidth may in return further degrade the depth accuracy. Consequently, it has become more and more challenging to engineer a high-performing lock-in pixel.

In this work, we propose and demonstrate novel lock-in pixels for indirect TOF based 3D sensing by the use of germanium (Ge)-on-Si technology. Ge-on-Si technology [14] has been widely adopted for making optical receivers on Si photonics platform, especially for high-speed optical communication applications. For examples, lateral p-i-n [15] and vertical p-i-n [16-19] photodiodes in normal-incidence configuration, as well as p-i-n [20-23] and metal-semiconductor-metal [24] photodiodes in waveguide configuration, have been previously demonstrated. Due to the lower bandgap of Ge compared to Si, Ge features a strong absorption at NIR wavelengths. When used in a lock-in pixel, such a property boosts the quantum efficiency and the frequency bandwidth, which in return reduces the depth error measured in an indirect TOF measurement. Moreover, Ge-on-Si technology is CMOS fabrication process compatible, which can be incorporated with the existing front-side illumination (FSI) and back-side illumination (BSI) CMOS image sensor processes in a straightforward manner.

The organization of this paper is arranged as below: in section 2, we layout a physical model to introduce the basic concepts of indirect TOF system. In section 3, we numerically simulate the electrical-optical properties of a proposed Ge-on-Si lock-in pixel with a finite-element approach. In section 4, we characterize the electrical-optical properties of the fabricated Ge-on-Si lock-in pixel with direct-current (dc) and alternating-current (ac) measurements. Based on these results, in section 5, we compare the system performance using either a conventional Si lock-in pixel or the proposed Ge-on-Si lock-in pixel by applying the physical model introduced in section 1. In particular, we analyze the issue of laser safety at possible operation wavelengths in section 6. Finally, a summary is given in section 7.

## 2. Physical model

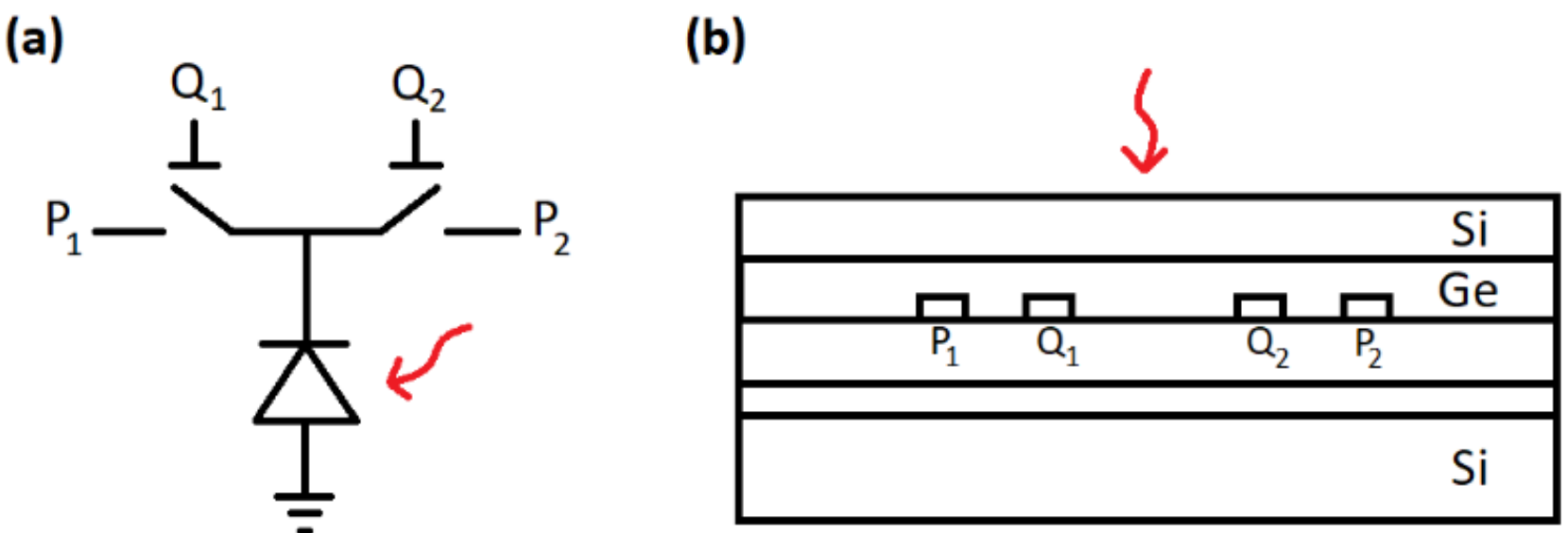

Fig. 1. (a) The equivalent circuit of a two-tap lock-in pixel. (b) The cross-sectional view of a Ge-on-Si lock-in pixel in a BSI configuration.

Consider the equivalent circuit of a two-tap lock-in pixel with $P_1$-$Q_1$-$Q_2$-$P_2$ configuration as shown in Fig. 1(a). The input optical power of the laser light (red wiggly arrow) is sinusoidally modulated at operation frequency $f_m$ with $\varphi_o$ initial phase; the photo-generated electrical current from the photodiode (black diode symbol) is sinusoidally demodulated by the voltages applied to the two Q nodes of the two switches (black switch symbol) at operation frequency $f_m$ with $\varphi_e$ (for $Q_1$) and $\varphi_e+\pi$ (for $Q_2$) initial phases. Then, the output electrical currents collected at the two P nodes can be expressed as

$$I_{\pm}(t)=\left[\frac{1\pm C_D^{\mathrm{dc}}\cos(2\pi f_m t-\varphi_e)}{2}\right]\left[I_a^{\mathrm{dc}}+I_p^{\mathrm{dc}}+I_p^{\mathrm{ac}}(f)\cos(2\pi f_m t-\varphi_o)\right]+I_d^{\mathrm{dc}}\,. \tag{1}$$

$C_D^{\mathrm{dc}}$ is the dc demodulation contrast; $I_a^{\mathrm{dc}}$ and $I_d^{\mathrm{dc}}$ are the ambient and dark currents; $I_p^{\mathrm{dc}}$ and $I_p^{\mathrm{ac}}(f)$ are the dc and ac photo currents in which $I_p^{\mathrm{ac}}(0)=I_p^{\mathrm{dc}}$. The ± sign represents the output electrical current collected at $P_1$ node or $P_2$ node. Physically, the voltages applied to the two Q nodes establish the designated electrical connection to the two P nodes instantaneously, representing the first bracket in (1), and it requires a transit time for the photo-generated electrical current of the photodiode to be collected by the two P nodes, representing the second bracket in (1). The dark current, ambient current, and photo current are assumed to be caused by material defect, ambient light, and laser light, respectively. Now, the time-average output electrical currents collected at the two P nodes can be derived as

$$\langle I_{\pm}\rangle'=\frac{1}{2}I_p^{\mathrm{dc}}\pm\frac{1}{4}C_D^{\mathrm{dc}}I_p^{\mathrm{ac}}(f)\cos(\varphi_o-\varphi_e)\,, \tag{2}$$

in which the photo current component is included but the ambient and dark current components are excluded. If we define an "ac" demodulation contrast as

$$C_D^{\mathrm{ac}}(f) \triangleq \left. \frac{\langle I_+ \rangle' - \langle I_- \rangle'}{\langle I_+ \rangle' + \langle I_- \rangle'} \right|_{\max \text{ e.g. } |\varphi_o - \varphi_e| = 0} = \frac{C_D^{\mathrm{dc}}}{2} \frac{I_p^{\mathrm{ac}}(f)}{I_p^{\mathrm{dc}}} \tag{3}$$

and insert it into (2), we arrive at

$$\langle I_\pm \rangle' = \frac{1}{2} I_p^{\mathrm{dc}} \pm \frac{1}{2} C_D^{\mathrm{ac}}(f) I_p^{\mathrm{dc}} \cos(\varphi_o - \varphi_e)\ . \tag{4}$$

By comparing (2) and (4), the concept of defining an "ac" demodulation contrast in (3) becomes clear, i.e., in the time-average sense, the transit time induced frequency response originally embedded in the second bracket in (1) is now effectively embedded in the first bracket in (1). Note that unlike the range of $C_D^{\mathrm{dc}}$ that is between 0~1, the range of $C_D^{\mathrm{ac}}(f)$ is between 0~0.5 due to the use of sinusoidal modulation and demodulation.

The stored charges on the two floating-diffusion capacitors connected to the two P nodes over duration $\tau$ can be then calculated by

$$\begin{aligned} qN_\pm(\varphi_e, \varphi_o) &= \lim_{\tau \gg \frac{1}{f_m}} \int_0^\tau I_\pm(t)dt = \frac{1}{2}(2I_d^{\mathrm{dc}} + I_a^{\mathrm{dc}})\tau + \frac{1}{2} I_p^{\mathrm{dc}} \tau \pm \frac{1}{2} C_D^{\mathrm{ac}}(f) I_p^{\mathrm{dc}} \tau \cos(\varphi_o - \varphi_e) \\ &\triangleq \frac{1}{2} qN_{dap} \pm \frac{1}{2} qN_p \cos(\varphi_o - \varphi_e) \end{aligned}\ . \tag{5}$$

By defining

$$\begin{aligned} qN_i &\triangleq qN_+(0, \varphi) - qN_-(0, \varphi) \\ qN_q &\triangleq qN_+(\frac{\pi}{2}, \varphi) - qN_-(\frac{\pi}{2}, \varphi) \end{aligned}\ , \tag{6}$$

it can be shown that the depth measured in the indirect TOF system can be expressed as

$$z = d_u \frac{\varphi}{2\pi} = \frac{c}{4\pi f_m} \tan^{-1}(\frac{N_q}{N_i})\ , \tag{7}$$

and the error of the depth measured in the indirect TOF system can be expressed as

$$\Delta z = d_u \frac{\Delta\varphi}{2\pi} = \frac{c}{4\pi f_m} \frac{\sqrt{N_{dap}}}{N_p}\ , \tag{8}$$

assuming the stored charges feature a Poisson distribution. $d_u$ is the unambiguous range and is equal to $c/(2f_m)$. The electron numbers associated with the dark current, ambient current, and photo current are $2I_d^{\mathrm{dc}}\tau/q$ (due to material defect), $I_a^{\mathrm{dc}}\tau/q$ (due to ambient light), and $I_p^{\mathrm{dc}}\tau/q$ (due to laser light), respectively, and will be denoted as $C$, $B$, and $A$ in the following. Finally, we can re-write (8) as

$$\Delta z = \frac{c}{4\pi f_m} \frac{1}{C_D^{\mathrm{ac}}(f_m)} \frac{\sqrt{A+B+C}}{A}\ . \tag{9}$$

It can be seen from (9) that to have a small depth error, the quantum efficiency that is linearly proportional to $A$ should be increased, the operation frequency should be increased, and the frequency bandwidth of the ac demodulation contrast should be increased to keep a high ac demodulation contrast at a high operation frequency. As will be elaborated in the next section, these properties make the proposed Ge-on-Si pixel a strong contender against the conventional Si pixel.

## 3. Numerical simulation

We apply a two-dimensional (2D) finite-element approach to simulate the steady and transient states of the proposed Ge-on-Si pixel. The cross-sectional view of the device in a BSI configuration is shown in Fig. 1(b). The layers from top to bottom are top Si layer, Ge layer, metal/oxide interconnect layer for pixels, metal/oxide interconnect layer for circuits, and bottom Si layer. The top three layers are from a donor wafer (pixel metal/oxide interconnect on Ge on Si on the donor wafer), and the bottom two layers are from a carrier wafer (circuit metal/oxide interconnect on Si on the carrier wafer). These two wafers are bonded at the wafer bonding interface in between the metal/oxide interconnect layer for pixels and the metal/oxide interconnect layer for circuits, and then the top Si layer is thinned down to a few microns. An anti-reflection coating (ARC) is applied to the backside of the top Si layer, so that the laser light with a finite size (to mimic the effect of an optical aperture over the backside of the top Si layer and in between the two Q nodes) injected from the backside of the top Si layer can be absorbed by the Ge layer efficiently. The two-tap lock-in pixel with $P_1$-$Q_1$-$Q_2$-$P_2$ configuration is implemented on the surface of Ge on Si on the donor wafer. As previously described, voltages applied to Q1 and Q2 nodes of the two switches result in electrical "on-off" or "off-on" state, and steer the photo-generated electrical current to be collected ideally by either P1 or P2 node. In the simulations, the Ge electron and hole mobilities are taken as 3900 and 1800 $cm^2$/V/s, respectively. Also, additional dopants are introduced in Ge to mimic the lattice-mismatch induced charge defects arisen from growing Ge on Si substrate. Moreover, we apply a trap-assisted bulk Shockley-Read-Hall model and a surface recombination model to match the experimentally measured dark currents of the vertical p-i-n photodiode in Ref. [17]. All simulations are done at 300 K temperature.

In Fig. 1(a), the simulated quantum efficiencies at the two P nodes are plotted as a function of the differential voltage applied to the two Q nodes. It can be seen that when the voltage applied to $Q_1$ is higher than that applied to $Q_2$, photo-generated carriers tend to be collected by $P_1$, i.e., the two effective switches are in the "on-off" state, and vice-versa. The total quantum efficiency, which is the sum of the quantum efficiencies at the two P nodes, is as high as ~ 93 % regardless of the differential voltage bias condition. In Fig. 1(b), the simulated dc demodulation contrast is plotted as a function of the differential voltage applied to the two Q nodes. It can be seen that the dc demodulation contrast increases first linearly and then sub-linearly with the differential voltage applied. In Fig. 1(c), the simulated small-signal frequency response of the ac demodulation contrast is plotted as a function of frequency. Instead of applying a differential voltage varying sinusoidally as a function of time to the two Q nodes and detecting the low-frequency signal amplitudes at the two P nodes, here we apply Eqn. (4) to calculate the small-signal response, i.e., by preparing an input optical delta or step function, monitoring the "smeared" output electrical delta or step function, and then performing a fast Fourier transform of the "smeared" output electrical delta or step function with an appropriate re-normalization. It can be seen that the 0.5 dB and 3 dB frequency bandwidths of the proposed Ge-on-Si pixel are as large as ~412 MHz and >1GHz, respectively.

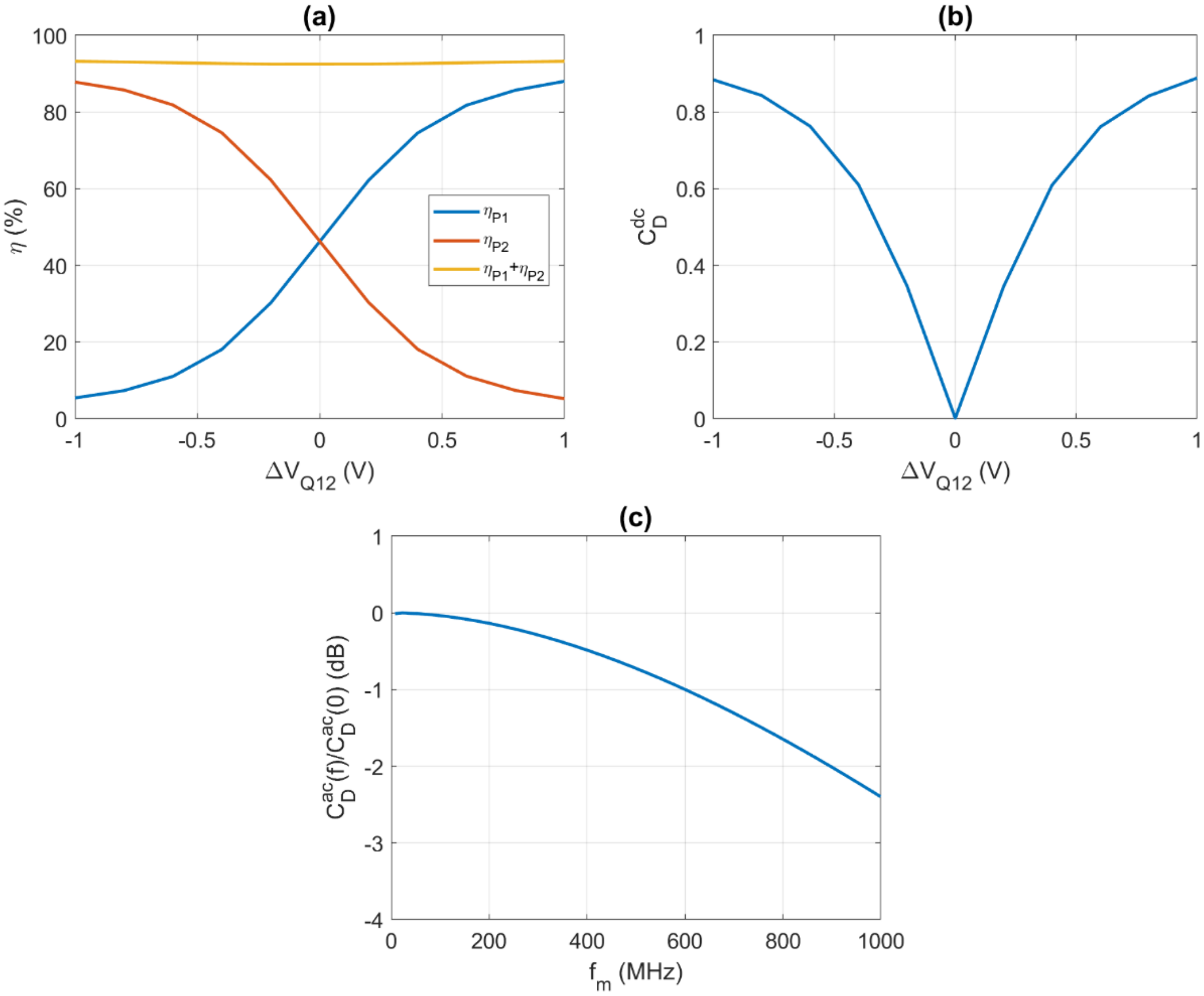

Fig. 2. The simulated (a) quantum efficiencies and (b) dc demodulation contrast plotted as a function of differential voltage. The simulated (c) normalized small-signal ac demodulation contrast plotted as a function of frequency. All simulations are performed with 940 nm laser light.

We have shown in Fig. 2 that at 940 nm the proposed Ge-on-Si pixel features a high quantum efficiency and a large frequency bandwidth compared to the conventional Si pixels [5-13]. Moreover, the proposed Ge-on-Si pixel can be operated at longer NIR wavelengths that mitigate the risk of human eye/skin damages. This is of particular importance for consumer electronics applications when a high depth accuracy is needed, e.g., indoor/outdoor navigations that typically requires a laser peak power of a few watts, since human eye may be irreversibly damaged when it is exposed to such a laser peak power at 850/940 nm wavelengths. In addition, shifting the operation wavelengths from 850/940 nm to longer NIR wavelengths can drastically increase the depth accuracy, since a much higher laser peak power can be applied yet still in compliance with laser-safety regulations. We will elaborate this part later in section 6. Note that while pixels made from some III-V semiconductor or organic material may also detect longer NIR wavelengths, the proposed Ge-on-Si pixel can benefit from its compatibility with CMOS foundry process, which eases the complexity in integrating electronic circuits, improves the fabrication yields, and drives down the chip cost. In Fig. 3, we re-simulate the quantum efficiencies, dc demodulation contrast, and ac demodulation contrast at 1550 nm wavelength as an example when using an eye-safe, or, strictly speaking, retina-safe laser (wavelengths > 1.4 μm). Note that the Ge-on-Si tensile strain induced enhancement of absorption coefficient at 1550 nm is assumed to be 3460 $cm^{-1}$ [16]. In Fig. 3(a), the total quantum efficiency is as high as ~ 43 % regardless of the differential voltage bias condition. Interestingly, in Fig. 3(b) and 3(c), we observed very similar results compared to Fig. 2(b) and 2(c). This is somewhat surprising at the first sight as the photo-generated carriers feature an exponential distribution and a almost constant distribution along the laser light incident direction, respectively, and should manifest different dc as well as ac demodulation contrasts, respectively. Nevertheless, we observe through transient simulations that the photo-generated carriers first diffuse and relax the original carrier distributions at the time scale of sub ns, and then experience the demodulation at a time scale > 1 ns for operation frequencies up to 1 GHz, which explains the similarities between Fig. 3(b)~(c) and Fig. 2(b)~(c).

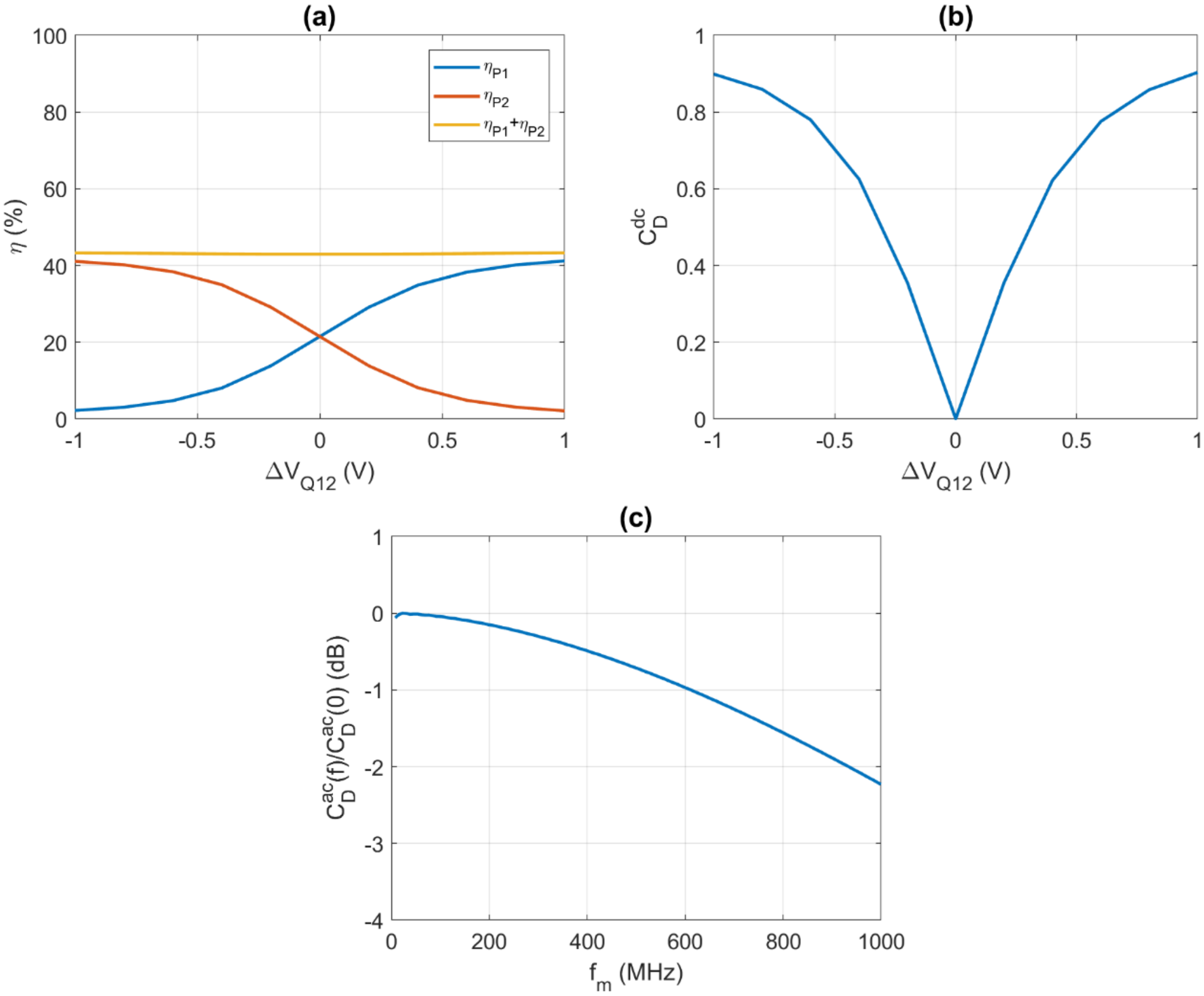

Fig. 3. The simulated (a) quantum efficiencies and (b) dc demodulation contrast plotted as a function of differential voltage. The simulated (c) normalized small-signal ac demodulation contrast plotted as a function of frequency. All simulations are performed with a 1550 nm laser light.

## 4. Device fabrication and characterization

The device is fabricated by first an epitaxial growth of Ge on Si on the donor wafer, and then going through the front-end pixel formation, back-end pixel metallization, and finally wafer bonding process along with the carrier wafer. For dc measurements, the fabricated Ge-on-Si pixels are characterized via a collimator coupled to a wavelength-stabilized 940/1550 nm laser source. For ac measurements, a pulse pattern generator is used to apply a modulated signal and a differential signal to an electro-optical modulator for modulating the input optical power and to a Ge-on-Si pixel for demodulating the photo-generated electrical current, respectively.

In Fig. 4(a), we plot the measured as well as the simulated quantum efficiencies at 940 nm as a function of the differential voltage applied to the two Q nodes. Consider the fact that in the simulation we use a nearly perfect ARC of ~ 99 % transmission but in the experiment an ARC of ~ 93 % transmission is used, the corresponding absorptions of Ge are estimated to be 94 % and 91 %, respectively. Such a discrepancy may be related to the mismatch between the simulated size of the laser light and the fabricated size of the optical aperture. Moreover, we observe the fabricated switch turns on more sharply (as a function of the differential voltage) in the experiment compared to the simulation. This may be related to the actual laser light profile is Gaussian in the experiment rather than uniform as in the simulation, and consequently generates more photo-carriers at the pixel center that effectively renders a stronger demodulation between the two Q nodes. In Fig. 4(b), the measured dc demodulation contrast at 940nm is shown to be higher than the simulated dc demodulation contrast, a manifestation of the fact that the fabricated switch turns on more sharply in the experiment compared to the simulation as shown in Fig. 4(a). In Fig. 4(c), the measured ac demodulation contrast at 940nm is shown to be slightly better than the simulated ac demodulation contrast but it is within the measurement error. Note that the measured ac demodulation contrast data points are normalized to the first data point, i.e., at 50 MHz, when they are plotted in Fig. 4(c). We repeat all these measurements again at 1550nm and the results are shown in Fig. 5. Similar traits as in Fig. 4 are observed, such as a lower absorption of Ge, a higher dc demodulation contrast, and a similar ac demodulation in the experiment compared to the simulation.

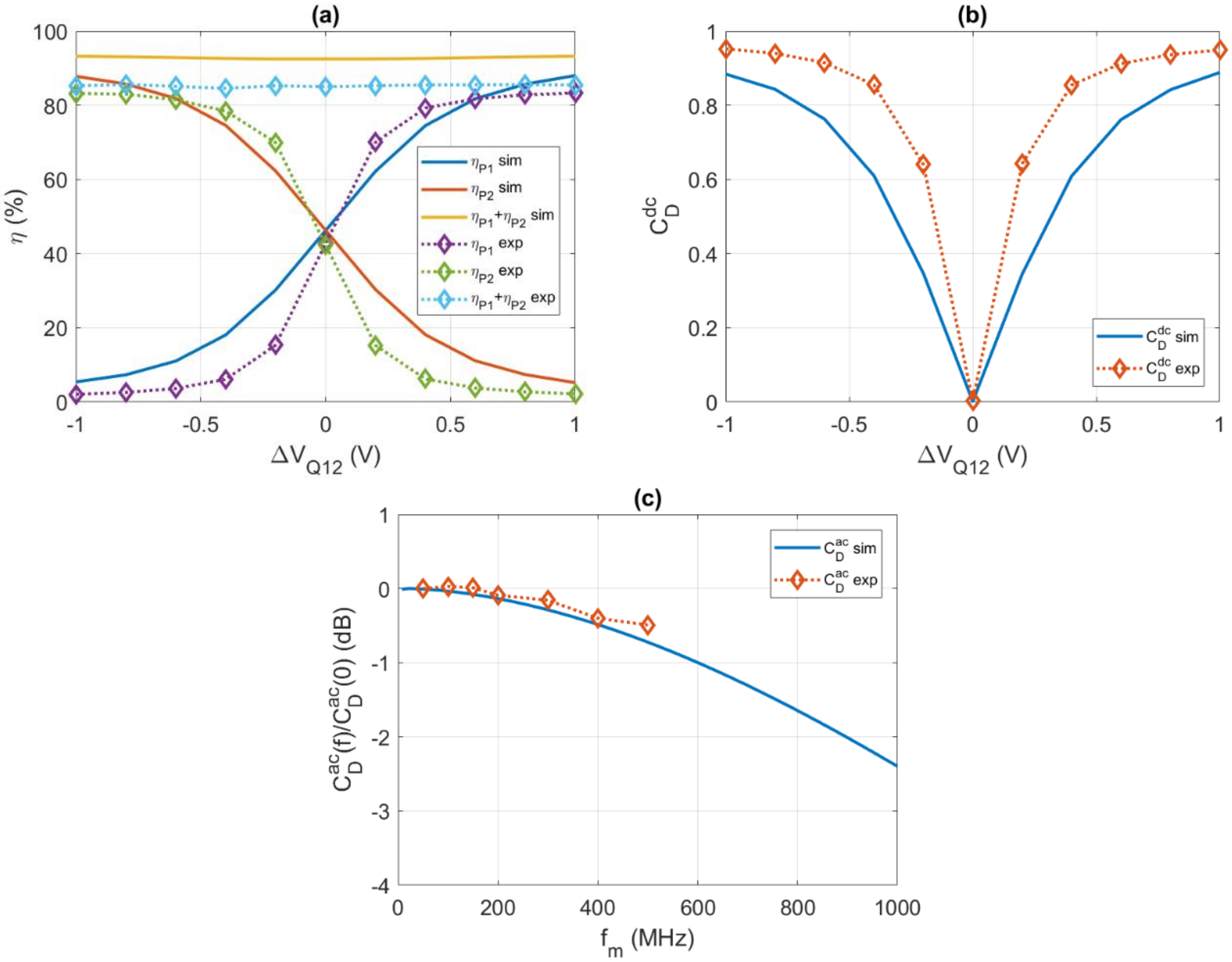

Fig. 4. The simulated and measured (a) quantum efficiencies and (b) dc demodulation contrast plotted as a function of differential voltage. The simulated and measured (c) normalized small-signal ac demodulation contrast plotted as a function of frequency. All simulations and measurements are performed with 940 nm laser light.

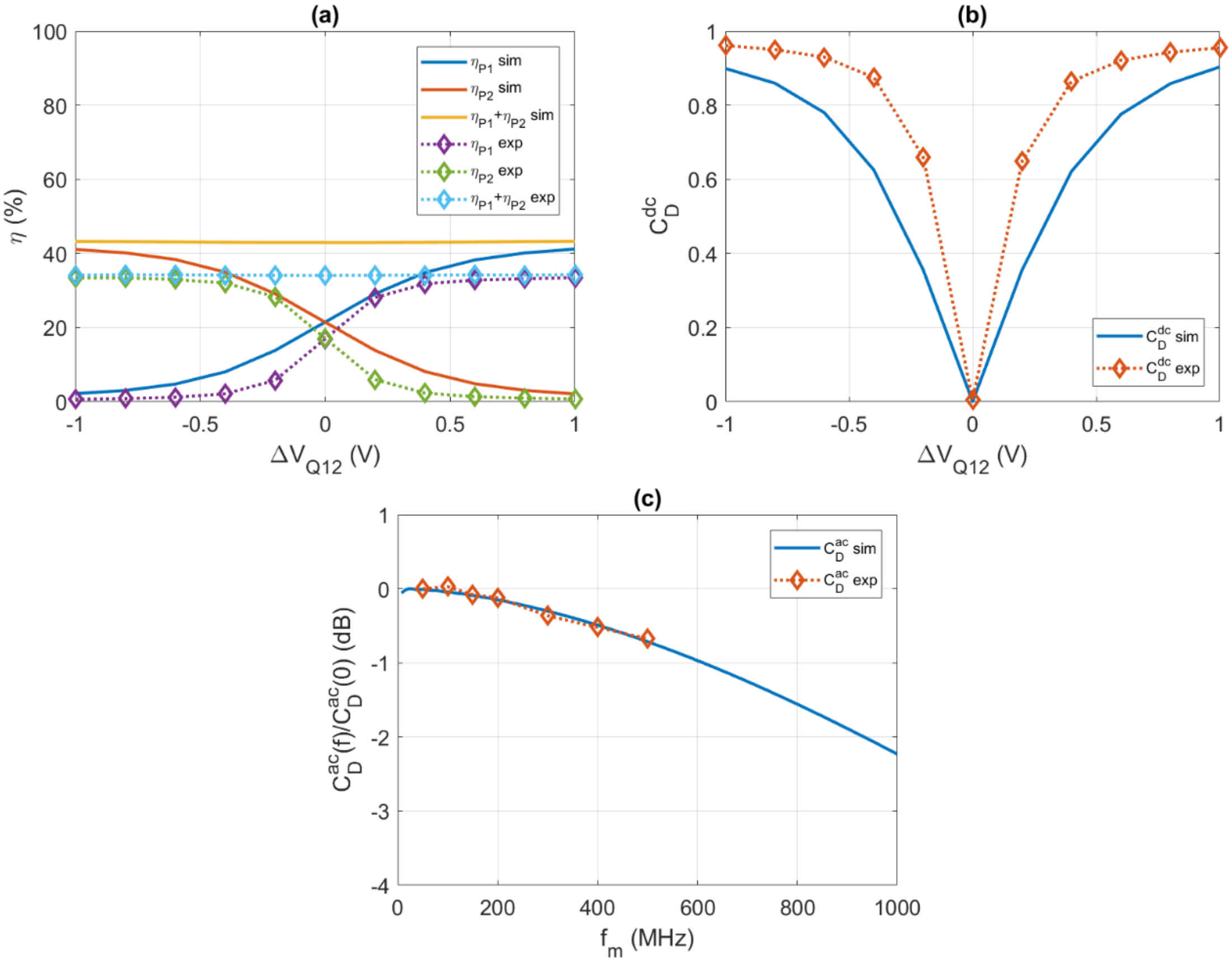

Fig. 5. The simulated and measured (a) quantum efficiencies and (b) dc demodulation contrast plotted as a function of differential voltage. The simulated and measured (c) normalized small-signal ac demodulation contrast plotted as a function of frequency. All simulations and measurements are performed with 1550 nm laser light.

## 5. System performance comparison

In the following, we calculate and compare the depth errors at 940 nm between the proposed Ge-on-Si pixel and the conventional Si pixel. For the Ge-on-Si pixel, the operation frequency, ac demodulation contrast, and quantum efficiency are given based on the simulation of an optimized device, in which a slightly lower quantum efficiency is traded for a slightly higher ac demodulation contrast. For the Si pixel, the operation frequency and ac demodulation contrast are estimated based on the experimental data shown in Ref. [13], and the quantum efficiency is estimated based on Ref. [25]. These parameters are listed in Table 1. The dark currents are estimated based on the experimental data shown in Ref. [15-24]. The outdoor ambient light spectral intensity is given based on the air mass (AM) 1.5 condition, and the ambient light spectral intensity of the indoor case is assumed to be 100 times smaller than that of the outdoor case.

In Fig. 6(a), we plot the depth errors of the Ge-on-Si and Si pixels as a function of depth, for both the indoor and outdoor cases, given laser power equal to 2 W. It is observed that, for the indoor case, the depth error of the Ge-on-Si pixel is consistently better than that of the Si pixel; for the outdoor case, again, the depth error of the Ge-on-Si pixel is consistently better than that of the Si pixel but at a larger scale. Moreover, for the Ge-on-Si pixel, the depth error of the indoor case increases by only a fraction when moving to the outdoor case; for the Si pixel, the depth error of the indoor case increases by almost an order of magnitude when moving to the outdoor case. In Fig. 6(b), we plot the depth errors of the Ge-on-Si and Si pixels as a function of laser power, for both the indoor and outdoor cases, given depth equal to 1 m. Similar trends as in Fig. 6(a) are observed. These results might be surprising as the dark current of the Ge-on-Si pixel is set to be 3 orders of magnitude larger than that of the Si pixel (nearly no changes to Fig. 6(a) and 6(b) even if a lower Si pixel dark current is set). The reason lies in that, in an indirect TOF system, in additional to the enhanced system signal by the larger operation frequency, ac demodulation contrast, and quantum efficiency, the dominant system noise is in fact due to the ambient light and the laser light instead of the dark current for various indirect TOF based 3D sensing scenarios. This is in sharp contrast to the earlier efforts in replacing indium-gallium-arsenide (InGaAs) with Ge-on-Si to make sensors used in the short wavelength infrared (SWIR) camera [26], in which weak ambient light is detected without any active NIR lighting so that the system noise is limited by the pixel dark current, but analogous to the recent efforts in replacing III-V compound semiconductors with Ge-on-Si to make receivers used in high-speed optical communication [27], in which the system noise is limited by the input referred noise of transimpedance amplifier (TIA) instead of the detector dark current. Note that we have also studied the multi-frequency de-aliasing algorithm [8] to extend the unambiguous range, and found it has minimum impact on the depth errors shown in Fig. 6(a) and 6(b).

**TABLE 1**

Estimated parameters used for system performance comparison

| | Si Pixel | Ge-on-Si Pixel |
|---|---|---|
| Operation Wavelength | 940 nm | |
| Operation Frequency | 100 MHz | 300 MHz |
| ac Demodulation Contrast at Operation Frequency | 0.425 | 0.47 |
| Quantum Efficiency | 20 % | 90 % |

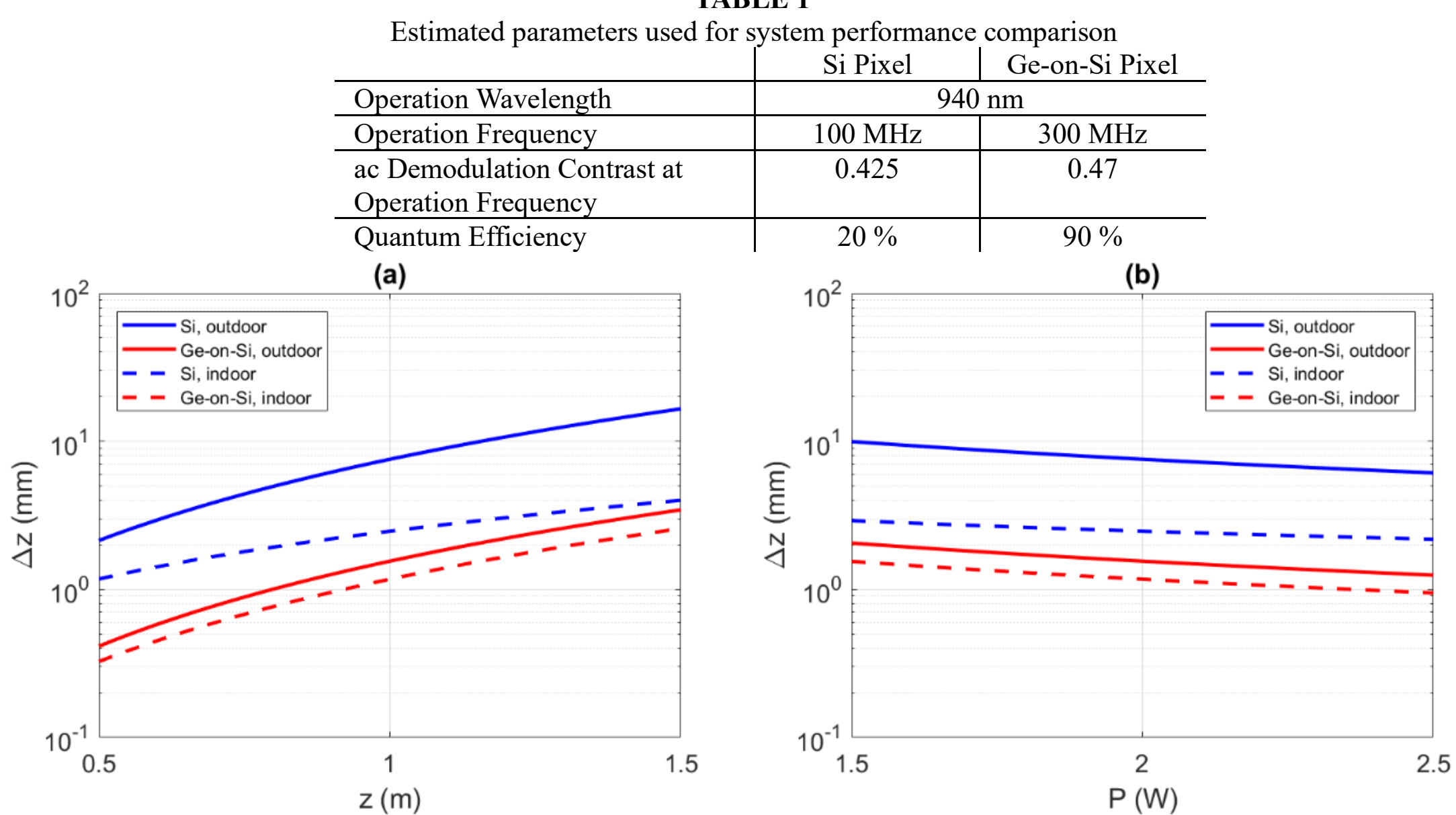

Fig. 6. (a) Depth errors plotted as a function of depth given laser power equal to 2 W. (b) Depth errors plotted as a function of laser power given depth equal to 1 m. Both indoor/outdoor conditions are considered.

## 6. Laser safety analysis

Since lasers operated at higher powers have been routinely used in indirect ToF 3D sensing to increase the system signal-to-noise ratio, laser safety should be carefully examined to ensure human eye/skin are free from laser-induced damages. Here we consider the international standard IEC 60825-1:2007 [28] and IEC 60825-1:2014 [29], and perform rigorous calculations to showcase a benefit of the Ge-on-Si lock-in pixel, i.e., it's possible to operate at longer NIR wavelengths that minimize or even avoid the risk of human eye/skin damages.

In these two international standards, the laser pulse energy is compared to the accessible emission level (AEL) to categorize the Class of a laser product. Class 1 refers to the laser product that is safe under all conditions of normal use. However, it should be noted that such a statement is only true for specified measurement conditions. E.g., in both IEC 60825-1:2007 regulation and IEC 60825-1:2014 regulation, the so-called "Condition 3" in Class 1 is to set the apparent source (laser emitter) away from the aperture stop (unaided eye) by 10 cm distance. Therefore, even if a laser product belongs to Class 1, it only guarantees under Condition 3 the eye is free from laser-induced damages for distances larger than 10 cm. The eye may still be harmed for distances smaller than 10 cm, especially for people with myopia. Such an issue may prevent wide adoption of 3D sensing technology in cell phone, tablets, and laptops etc. consumer electronics products.

In the following, we consider a typical setting of indirect ToF 3D sensing: the laser peak power ranges from 2.5 W to 5 W; the optical format follows a 4-quadrature-phase sequence with 500 μs optical exposure time and 100 μs electrical readout time; the frame rate is 60 fps. We then calculate the maximally allowed laser peak powers at specified measurement conditions as a function of wavelength, assuming the laser pulse energy is kept a constant fraction of AEL. It is shown in Fig. 7(a) and Fig. 7(b) that by moving the operation wavelength from 940 nm to 1350 nm (the wavelength in which the solar ambient light starts to diminish), the maximally allowed laser peak powers are enhanced by roughly 1 and 4 orders of magnitude following IEC 60825-1:2007 regulation and IEC 60825-1:2014 regulation, respectively. Note that the major difference between the two regulations from 1.3 to 1.4 μm wavelength is that the safety factor, defined as the effective dose at 50% probability producing a minimal visible lesion (MVL) divided by the maximum permissible exposure (MPE), is significantly increased in the newer version [30]. We also calculate the minimally allowed safe distance for different wavelengths at 100% AEL. It is found that for 940 nm, the minimally allowed safe distance between laser source and human eye is about 4 to 8 cm and 3 to 4 cm following IEC 60825-1:2007 regulation and IEC 60825-1:2014 regulation, respectively; for 1350 nm, it is about 6 to 9 mm and 0 mm following IEC 60825-1:2007 regulation and IEC 60825-1:2014 regulation, respectively. These results showcase the benefit of the Ge-on-Si lock-in pixel for indirect ToF 3D sensing, i.e., the flexibility to operate at a longer NIR wavelength ensures unprecedented safety of the users.

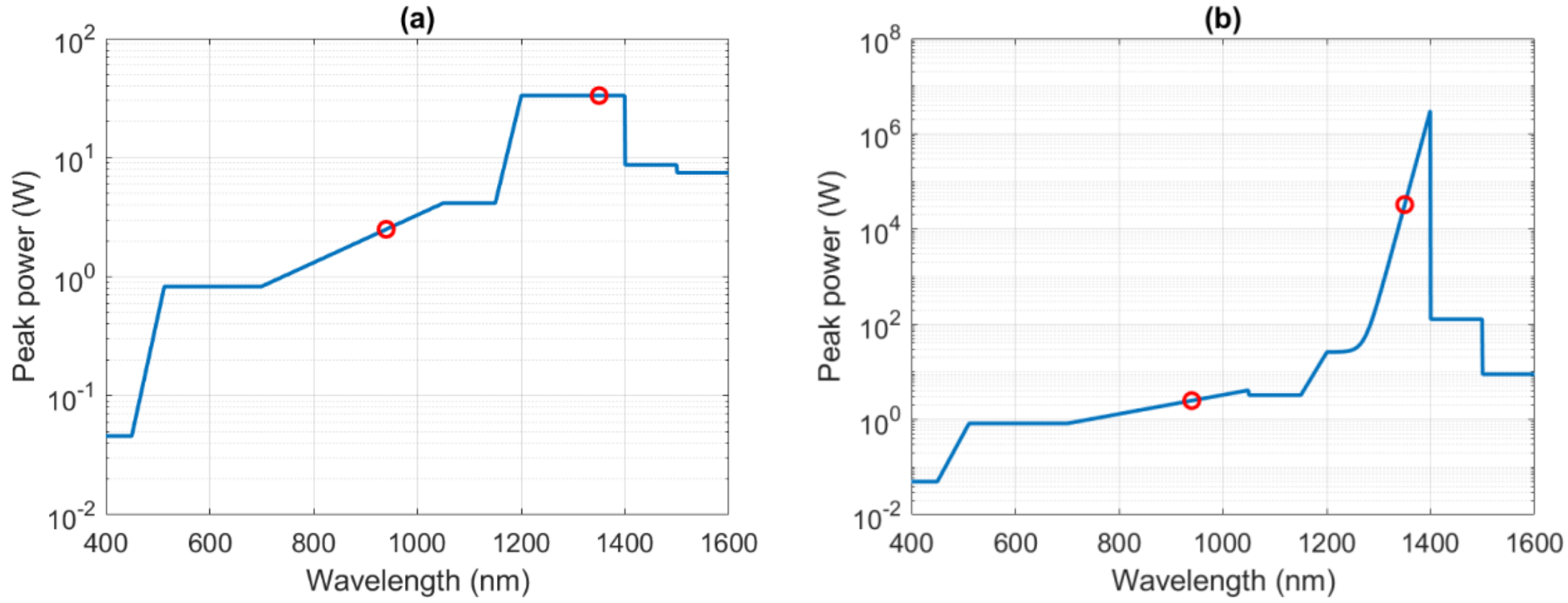

Fig. 7. (a) The maximally allowed laser peak power plotted as a function of wavelength considering IEC 60825-1:2007 regulation under Condition 2 (the strictest among Conditions 1, 2, and 3), in which the laser pulse energy is set to be a constant fraction of AEL. (b) The maximally allowed laser peak power plotted as a function of wavelength considering IEC 60825-1:2014 regulation under Condition 3 (the strictest among Conditions 1 and 3), in which the laser pulse energy is set to be a constant fraction of AEL.

## 7. Summary

We propose and demonstrate the use of Ge-on-Si technology to create a novel lock-in pixel for indirect TOF based 3D sensing applications. A high quantum efficiency and a large frequency bandwidth are simulated and measured at 940 nm as well as at 1550 nm wavelengths, which can be further optimized in the future. The system performance comparison between a Si pixel and a Ge-on-Si pixel at 940nm indicates that the Ge-on-Si pixel features lower depth errors compared to the Si pixel in both indoor and outdoor scenarios. The laser safety analysis showcases that when the Ge-on-Si pixel is operated at a longer NIR wavelength, it is substantially safer than at 940 nm wavelength, allowing the use of a much higher laser power but still in compliance with laser-safety regulations.

## Acknowledgment

We acknowledge the support of device fabrication by Taiwan Semiconductor Manufacture Company (TSMC) in Hsinchu, Taiwan, ROC.